# Creation of Helical Dirac Fermions by Interfacing Two Gapped Systems of Ordinary Fermions

Z. F. Wang[1†], Meng-Yu Yao[2†], Wenmei Ming[1], Lin Miao[2], Fengfeng Zhu[2], Canhua Liu[2], C. L. Gao[2], Dong Qian[2*], Jin-Feng Jia[2] & Feng Liu[1*]

[1]Department of Materials Science and Engineering, University of Utah, Salt Lake City, Utah 84112, USA

[2]Key Laboratory of Artificial Structures and Quantum Control (Ministry of Education), Department of Physics, Shanghai Jiao Tong University, Shanghai 200240, China

[†]These authors contributed equally to this work.

[*]Correspondence and requests for materials should be addressed to D.Q. (email: dqian@sjtu.edu.cn ) and F.L. (email: fliu@eng.utah.edu).

Topological insulators (TIs) have attracted a great deal of recent interest as a new class of materials[1-9]. One unique property of TIs is its electronic structure characterized by the helical Dirac cone states residing in the middle of a bulk insulating gap. Charge carries in such Dirac states are helical Dirac fermions (HDFs), which, different from ordinary fermions, are massless relativistic particles with their spin locked to their momentum ideally in perpendicular direction. On one hand, the HDFs are expected to give rise to a range of fundamentally new physical phenomena, such as anomalous half-integer quantum Hall effect[10,11], Majorana Fermions[12] and fractionally charged quantum particles[13]. On the other hand, the spin transport of HDFs is protected against back scattering by time-reversal symmetry, which affords a promising paradigm of spintronic devices.

So far, the HDFs are considered to be an intrinsic property of TIs, which only exist in TIs with sufficient size. If the dimension of 3D TIs is reduced, the coupling between the surface states coupled with quantum confinement effect will open an energy gap, eliminating the HDF states, *i.e.*, the linear Dirac bands of HDFs revert to the parabolic conventional bands of normal fermions upon gap opening[14-16]. For example, when the thickness of $Bi_2Se_3$ (or $Bi_2Te_3$), two well-known 3D TIs, is reduced to less than six quintuple layers (QLs), a gap opens[14-16]. While the ultrathin Bi (111) films of less than four bilayers (BLs), a 2D TI system, is characterized with a finite gap[17-20]. Therefore, the common expectation is that the HDFs cannot survive in the gapped systems.

Here we report the discovery of extrinsic formation of HDFs by interfacing two gapped thin films: a single BL Bi (111) film grown on a single QL $Bi_2Se_3$ or $Bi_2Te_3$ substrate. Using density function theory (DFT) band calculations, we predict the existence of helical Dirac cone in such supposedly ordinary fermion systems and the prediction is directly confirmed by experiment. Further theoretical analysis shows that the extrinsically created helical Dirac cone consists of predominantly Bi bilayer states. It is induced by a giant Rashba effect of band splitting with a coupling constant up to ~4 eV·Å, resulting from the internal electric field due to interfacial charge transfer. Our findings provide a promising new method to manipulate the electronic and topological states by interface engineering.

**Results**

**Construction of two gapped systems.** Our idea of interfacing two gapped films of single BL Bi (111) film and single QL $Bi_2Se_3$ or $Bi_2Te_3$ is partly motivated by existing knowledge of growing Bi film on different substrates. It is known that when Bi grown on metal[21,22] or semiconductor substrates (*e.g.* Si, Ge)[23-25], an extrinsic helical Dirac surface state can be created by the substrate induced spin-splitting of Bi bands, *i.e.* the Rahsba effect. However, generally there exist strong interactions between the Bi and the substrate leading to strong hybridization of surface and bulk states, and additionally metal substrate is undesirable for device applications. Recent experiments have grown Bi (111) bilayer on TI $Bi_2Te_3$[26,27] and $Bi_2Se_3$, which also see extrinsic formation of the helical Dirac states from Bi. But the problem remains that there exists the intrinsic Dirac state and other bulk states of $Bi_2Te_3$ substrate that can hybridize with the extrinsic surface Dirac state of Bi BL[27]. Therefore, one needs to find a suitable gapped substrate for growing Bi that will avoid strong hybridization between the extrinsic Bi Dirac states and substrate states, so as to better isolate surface Dirac transport. In this regard, growing Bi bilayer on a gapped 1QL $Bi_2Te_3$ (or $Bi_2Se_3$) substrate versus on a semimetal thick (>6QL) QL $Bi_2Te_3$ (or $Bi_2Se_3$) substrate makes a significant difference. Further, experimentally growing high quality 1QL $Bi_2Te_3$ (or $Bi_2Se_3$) film turns out to be much harder than thick films, as we discuss below.

**DFT band structures.** A single BL Bi (111) film is predicted to be a gapped 2D TI[17-20]. Our calculations show that at the equilibrium substrate lattice constant of $Bi_2Se_3$ and $Bi_2Te_3$, the strained Bi (111) film has an energy gap of ~0.17 eV (Fig. 1a) and ~0.44 eV (Supplementary Fig. S1a), respectively. On the other hand, $Bi_2Se_3$ and $Bi_2Te_3$ are well known 3D TIs with a single helical Dirac cone below Fermi level. However, when their thickness is reduced to less than 6QL, a finite energy gap opens removing the Dirac cone. As shown in Fig. 1b and Supplementary Fig. S1b, we can clearly see an energy gap in 1QL $Bi_2Se_3$ (~0.5 eV) and $Bi_2Te_3$ (~0.3 eV), which are consistent with previous works[15,16]. After 1BL Bi (111) is deposited onto 1QL $Bi_2Se_3$ (or $Bi_2Te_3$), the band structures are found to change dramatically. Most surprisingly, the gap vanishes as a linear dispersive band emerges around the Γ point to form the Dirac cone, marked with "D" in Figs. 1c-d and Supplementary Figs. S1c-d. For the Bi/$Bi_2Se_3$ (Figs. 1c and 1d), the Dirac point locates at ~0.1 eV below the Femi level with a Fermi velocity ($V_F$) ~$5.3\times10^5$ m/s; for the Bi/$Bi_2Te_3$ (Supplementary Figs. S1c and S1d), the Dirac point locates at ~0.2 eV below the Femi level with a $V_F$~ $4.5\times10^5$ m/s. We note that the extrinsic Dirac point formed in our system is not perfect as the second derivative of $E(\mathbf{k})$ doesn't exactly vanish. However, the $E(\mathbf{k})$ is almost linear with spin-momentum locking property, having a very large Fermi velocity comparable to that in graphene or other TIs. Therefore, from charge and spin transport point of view, our system works equivalently well for spintronics device applications.

**ARPES band structures.** Our DFT predictions are directly confirmed by experiments, as shown in Fig. 2. A single BL Bi (111) is grown on the (111)-oriented 1QL $Bi_2Se_3$ substrates, which were grown first on graphene prepared on SiC wafer[28]. We found that growth of high quality ultrathin film of 1QL $Bi_2Se_3$ or $Bi_2Te_3$ turned out to be much harder than growth of thick film of multi-QLs. So far, we only attained high quality of 1QL $Bi_2Se_3$ film to our satisfaction, but not for 1QL $Bi_2Te_3$. The thickness and quality of the films are monitored by in situ reflection high-energy electron diffraction (RHEED) and scanning tunneling microscopy (STM). The electronic bands are measured directly by angle-resolved photoemission spectroscopy (ARPES). In Fig. 2a, an energy gap is observed for the bare 1QL $Bi_2Se_3$. Bi can grow epitaxial with a unit of BL along (111) orientation on $Bi_2Se_3$. After deposition of a single BL Bi (111), the spectra are dramatically changed and a Dirac cone clearly appears around the Γ point, as shown in Figs. 2b-c. The DFT bands calculated at the experimental graphene lattice constant (green dashed lines) are overlaid with the ARPES spectra along high symmetric directions and the agreement between the theory and experiment is very good. In order to match experimental data, the Fermi level of DFT bands is slightly shifted. The Dirac cone is isotropic. Figure 2d and 2e presents the high resolution ARPES spectra near the Dirac cone and corresponding momentum distribution curves. The linearly dispersive bands of the Dirac cone are clearly resolved. The Dirac point is about 0.07 eV below Fermi level estimated from the crossing point of the linearly dispersive bands (green dotted line). The $V_F$ of the Dirac cone is ~$6\times10^5$ m/s, which is the same as that of bare $Bi_2Se_3$.

**Extrinsic helical Dirac states.** We have also confirmed the helical properties of the newly formed extrinsic Dirac cones in the interfaced systems of a single BL Bi (111) and 1QL $Bi_2Se_3$ (or $Bi_2Te_3$), which are individually gapped materials. HDFs are characterized by the spin-momentum locking relations[29]: (i) the spin direction is perpendicular to the momentum direction; (ii) the spins have the opposite directions at the inverse momenta. We have calculated three spin components of the Bloch state in the Brillouin zone, as shown in Fig. 3. Figure 3a shows the zoom-in band structure around the Dirac cone for the Bi/$Bi_2Se_3$ system, indicating two equal-energy cuts (circles) above and below the Γ point with $k$=0.04 Å$^{-1}$ (radius of circle). The in-plane spin projections along the upper and lower circles are plotted in Figs. 3c and 3d, respectively. The spins point clockwise around the upper circle, but counterclockwise around the lower circle, and they are in opposite directions at the inverse momenta. In addition, the spins are not completely in-plane, and their out-of-plane projections are plotted in Fig. 3e, which oscillate periodically around circle with a period of 120º (reflecting the threefold rotational symmetry) and have the opposite phase in the upper and lower circles. Thus, as the electrons trace the circle in the momentum space, their spins lie almost in the plane of the circle, but wobble above and below the plane periodically. The finite out-of-plane spin-component is due to the warping effect of the Dirac cone (see Fig. 3a), which has not a perfect linear dispersion. The six nodal points in the out-of-plane wobbling motion (Fig. 3e) correspond to high-symmetry Γ-M directions. We also calculated the angle ($α$) between the electron spin and momentum direction, as shown in Fig. 3f. $α$ is very small, less than 0.3º and 0.1º in the upper and lower circle, respectively, indicating that spins are almost perpendicular to their momenta. All these features demonstrate the spin-momentum locking property of the helical Dirac states. The same results and conclusions are drawn for the Bi/$Bi_2Te_3$ system, as shown in Supplementary Fig. S2.

**Discussion**

Next, we explain where the Dirac cone states come from, the 1BL Bi or 1QL $Bi_2Se_3$ (or $Bi_2Te_3$)? Previous studies[18] showed that the Bi (111) film represents a special class of system having an intermediate inter-bilayer coupling strength, which has a significant influence on its topological property. At equilibrium, the interfacial distance ($d$) between Bi BL and $Bi_2Se_3$ QL is ~3 Å and the interface energy is ~0.16 eV per unit cell, which are noticeably larger than the typical values of van der Waals bond but smaller than those of typical chemical bond. To gain some insight about the influence of the interfacial coupling on the helical spin states, we have performed a set of model DFT calculations by artificially increasing the interfacial distance between the Bi BL and $Bi_2Se_3$ QL to gradually tune the interface coupling strength[18], as shown in Fig. 4. At a large interfacial distance (e.g., $d$~6 Å in Fig. 4a), without the interfacial coupling, the band structure would be a simple superposition of the bare Bi BL and $Bi_2Se_3$ QL bands. However, we observe a slight band splitting in the double-degenerated Bi and $Bi_2Se_3$ band. With the decreasing interfacial distance, the increased interfacial coupling continues to further split and reshape the bands, leading eventually to the formation of helical Dirac cone. By analyzing the spectral components of Dirac cone states during this process, we found that the Dirac cone is predominantly coming from the Bi BL, with ~95% spectral weight of Bi BL states at the equilibrium distance, as shown in Fig. 4c. Similar results are found for the interface between Bi BL and $Bi_2Te_3$ QL, with ~90% Bi BL states in the helical Dirac cone at the equilibrium distance, as shown in Supplementary Figs. S3a-c.

Lastly, we will address how the Dirac cone states are formed. From the above calculations and analyses, we learnt that the extrinsic HDFs originate from the Bi BL, while the QL of $Bi_2Se_3$ (or Bi$_2$Te$_3$) substrate provides possibly a triggering mechanism that induces the Dirac cone states in Bi BL. We also know that Bi (111) BL has a strong spin-orbit coupling, and by itself its bands are double-degenerate conserving the inversion symmetry. Deposition of Bi BL on Bi$_2$Se$_3$ (or Bi$_2$Te$_3$) will break such symmetry, and hence lift the band degeneracy. However, this effect of symmetry breaking is limited, likely insufficient to induce a band splitting and reshaping as dramatic as we have observed. To further understand the interaction between the Bi BL and Bi$_2$Se$_3$ (or Bi$_2$Te$_3$), we calculated the differential charge density at the interface, defined as $\rho_{diff}=\rho_{BL+QL}-\rho_{BL}-\rho_{QL}$, as shown in Fig. 4d (see also Supplementary Fig. S3d for Bi/Bi$_2$Te$_3$). Clearly, a substantial charge transfer at the interface has occurred. The Bi BL acts as a donor and the Bi$_2$Se$_3$ (or Bi$_2$Te$_3$) as an acceptor, with the electrons transferring from the former to the latter. This charge transfer generates a large internal electric field at the interface region, with a field direction pointing from the Bi BL to the Bi$_2$Se$_3$ (or Bi$_2$Te$_3$), and field intensity estimated as high as ~1 V/Å. Thus, we propose this phenomenon is very similar to the external electric field (or band bending) effect that induces Rashba band splitting in the surface state of TIs[30-32]. We estimated the Rashba coupling constant (using the 2D free electron gas model) from the calculated band splitting at the equilibrium structures, which is ~4 eV·Å. Therefore, the HDFs are extrinsically created by a giant Rashba effect due to the interfacial charge transfer.

To further support the charge transfer induced Rashba effect is the dominated factor for realizing the extrinsic helical Dirac states, we also studied the band structure for Bi BL on 1QL In$_2$Se$_3$ and 5-layer Cu (111) substrate, as shown in Supplementary Fig. S4 and S5, respectively. A clearly charge transfer between the Bi BL and these non-TI substrates (Supplementary Figs. S4i and S5i) can be observed in both cases (one semiconductor and one metal), and the similar Dirac cone states are also generated. However, the Bi BL become metallic at the $In_2Se_3$ lattice constant, and the Cu (111) substrate is also metallic. Thus, comparing to $Bi/Bi_2Se_3$ and $Bi/Bi_2Te_3$ system, these Dirac cone states are not within the gapped region of $Bi/In_2Se_3$ and Bi/Cu. The previous studies for Bi grown on Si (111) only showed the monomer or trimer Bi configuration on the substrate, which cannot maintain the intrinsic structure of the Bi BL.

**Methods**

**Calculation method.** DFT calculations for a single BL Bi (111) on a single QL $Bi_2Se_3$ and $Bi_2Te_3$ are carried out in the framework of the PBE-type generalized gradient approximation using VASP package[33]. The experimental bulk lattice constants of the substrate are used (*a* =4.138 Å for $Bi_2Se_3$ and *a*=4.386 Å for $Bi_2Te_3$), and the Bi BL is strained to match the substrate lattice constant. All calculations are performed with a plane-wave cutoff of 400 eV on an 11×11×1 Monkhorst-Pack *k*-point mesh, and the vacuum layer is over 20 Å thick to ensure decoupling between neighboring slabs. During structural relaxation, all the atoms are allowed to relax until the forces are smaller than 0.01 eV/Å. To reproduce the ARPES band for $Bi/Bi_2Se_3$ in Fig. 2, we use a larger lattice constant *a*=4.37 Å, which can match the graphene/SiC wafer in the experiment.

**Experiment method.** Flat $Bi_2Se_3$ single QL films is grown by molecular beam epitaxy method on graphene/SiC wafer and on highly ordered pyrolytic graphite substrates. Bi films are deposited on $Bi_2Se_3$ at 200 K and then annealed at room temperature. The thickness of $Bi_2Se_3$ and Bi films is monitored by RHEED oscillation and STM. The sample temperature is kept at 100 K during measurement. ARPES measurements are performed with in-lab He discharge lamp (He-I 21.2 eV). Energy resolution is better than 25 meV and angular resolution is better than 0.02 Å$^{-1}$.

**Author Contributions**

Z.F.W. carried out the theoretical calculations with assistance from W.M.M. and F.L.; M.Y.Y. did the experiments with assistance from L.M., F.F.Z., C.H.L., C.L.G., D.Q. and J.F.J.; F.L, D.Q., and J.F.J. were responsible for the overall direction. Z.F.W., F.L. and D.Q. wrote the paper.

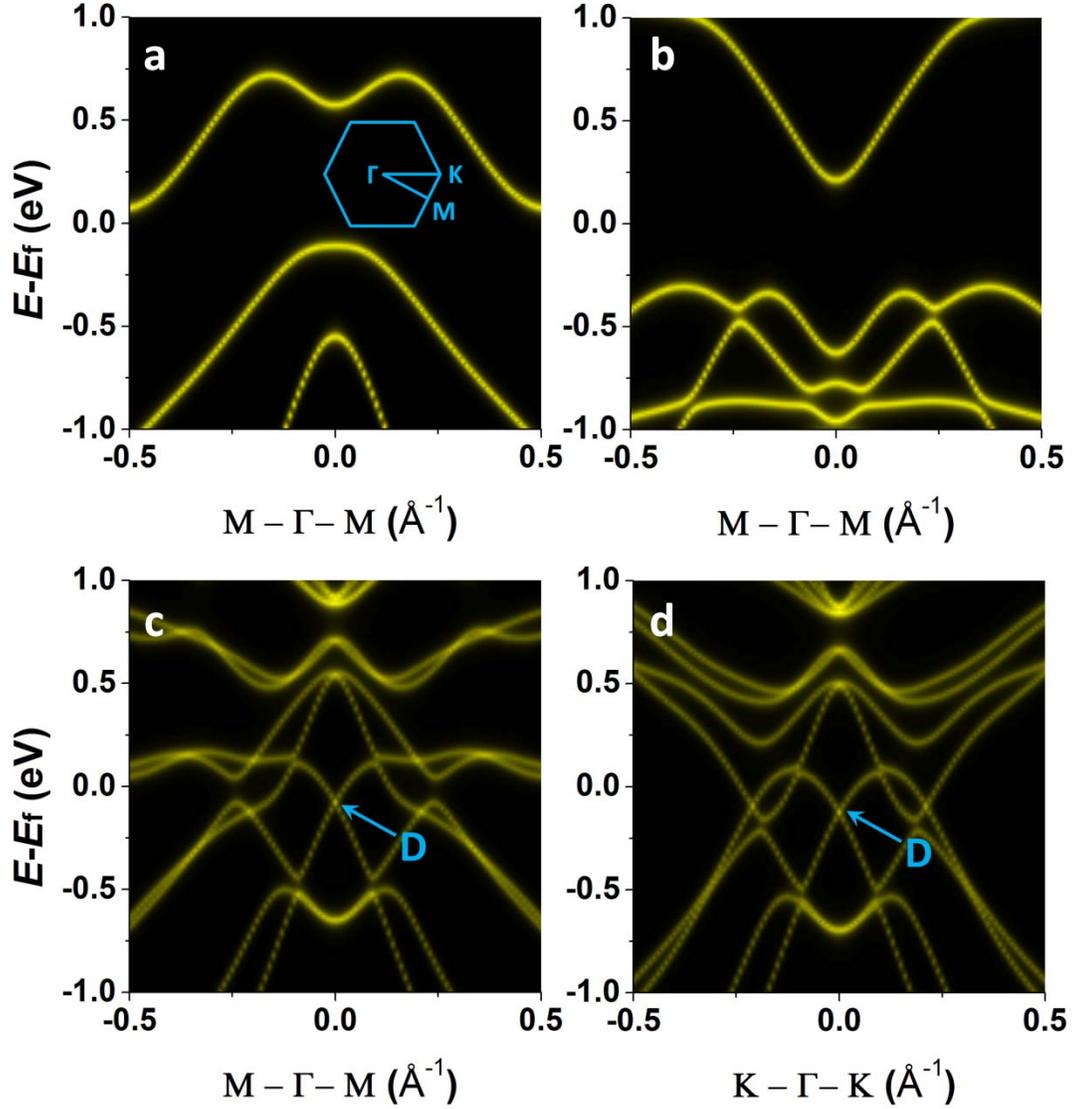

**Figure 1. Theoretical bands along the high-symmetry directions.** (a) Bare Bi (111) BL along M-Γ-M direction at the bulk $Bi_2Se_3$ lattice constant. The inset is the first Brillouin zone with high symmetric points of Γ, M and K. (b) Bare 1QL $Bi_2Se_3$ along M-Γ-M direction. (c) and (d) 1BL Bi (111) on 1QL $Bi_2Se_3$ along M-Γ-M and K-Γ-K directions, respectively. The newly created Dirac cone is marked with "D". The relative intensity of bands at different k-points scales with the number of local density of states.

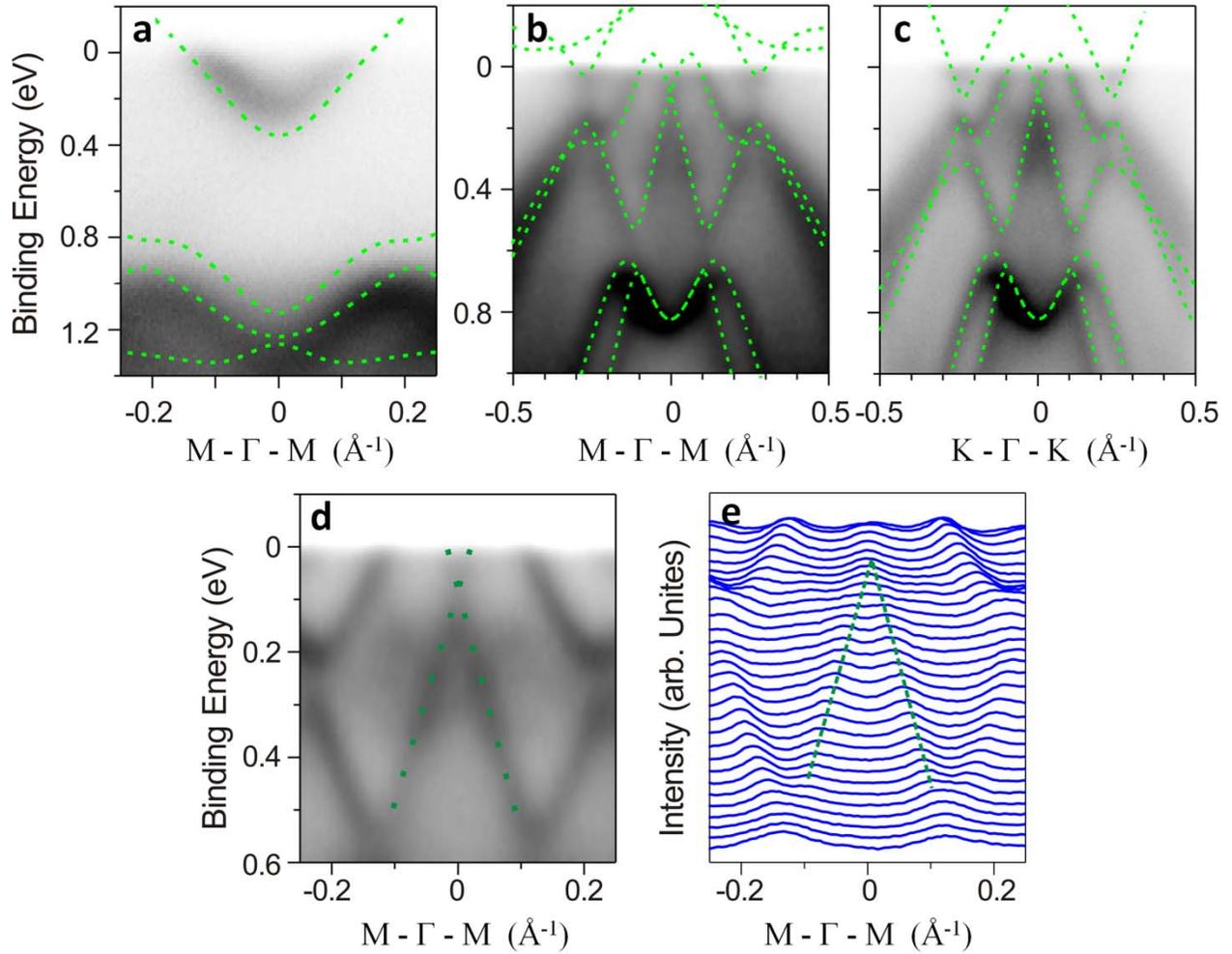

**Figure 2. Experimental bands superimposed with theoretical bands.** (a) Bare 1QL Bi$_2$Se$_3$. (b) and (c) 1BL Bi (111) on 1QL Bi$_2$Se$_3$ along M-Γ-M and K-Γ-K directions, respectively. Green dashed lines are the theoretical bands. (d) and (e) High resolution ARPES spectra near the Dirac cone and corresponding MDC curves. Green dotted line marks the linearly dispersive bands that crossing at Dirac point at ~70 meV blew Fermi level. Note that the calculated bands are slightly different from Fig. 1 because different lattice parameter used.

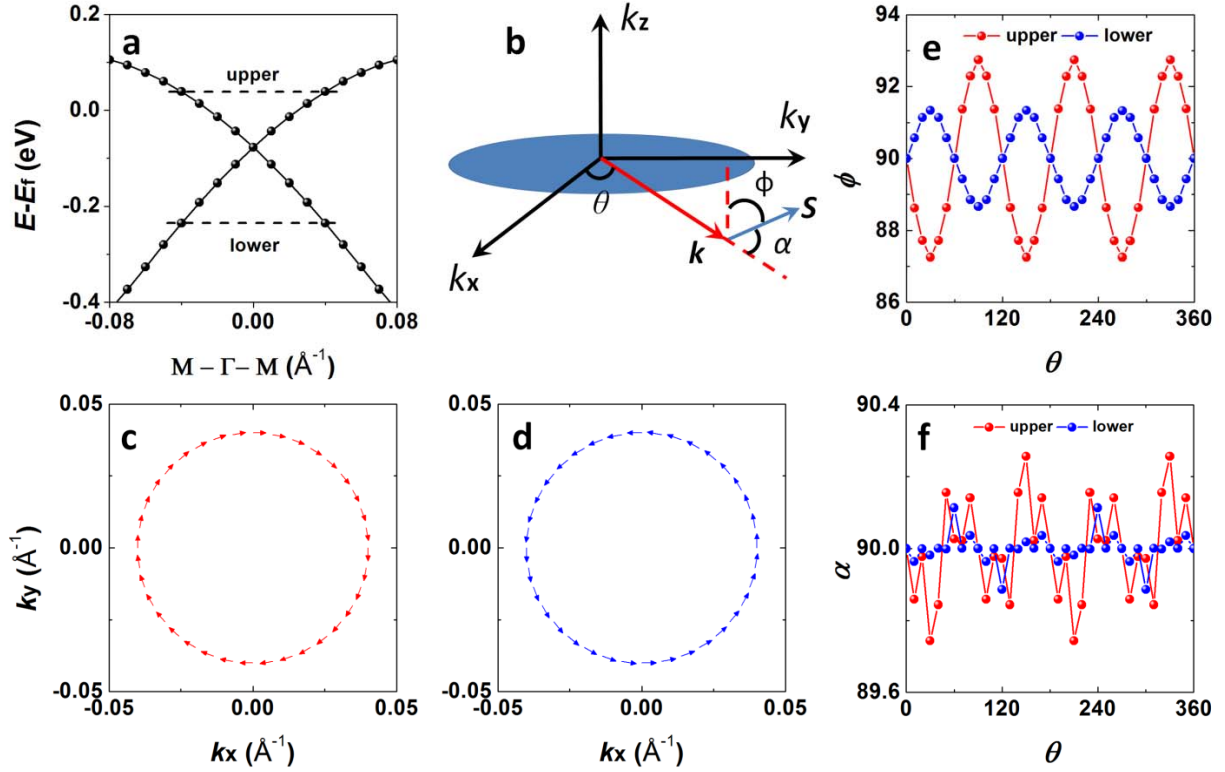

**Figure 3. Helical states near the Dirac cone.** (a) Two equal-energy cuts as marked above and below Dirac point for calculating the helical Dirac states in Bi/Bi$_2$Se$_3$. (b) Definitions of the $\theta$ (angle between **k** and $k_x$ axis), $\phi$ (angle between **S** and $k_z$ axis) and $\alpha$ (angle between **k** and **S**), which are used to characterize the spin direction (**S**) with respect to the momentum direction (**k**). (c) and (d) In-plane spin-momentum relation in the upper and lower cut in (a), respectively. (e) Out-of-plane spin-momentum relation in the upper and lower cuts in (a). (f) Angle between spin and momentum in the upper and lower cuts in (a).

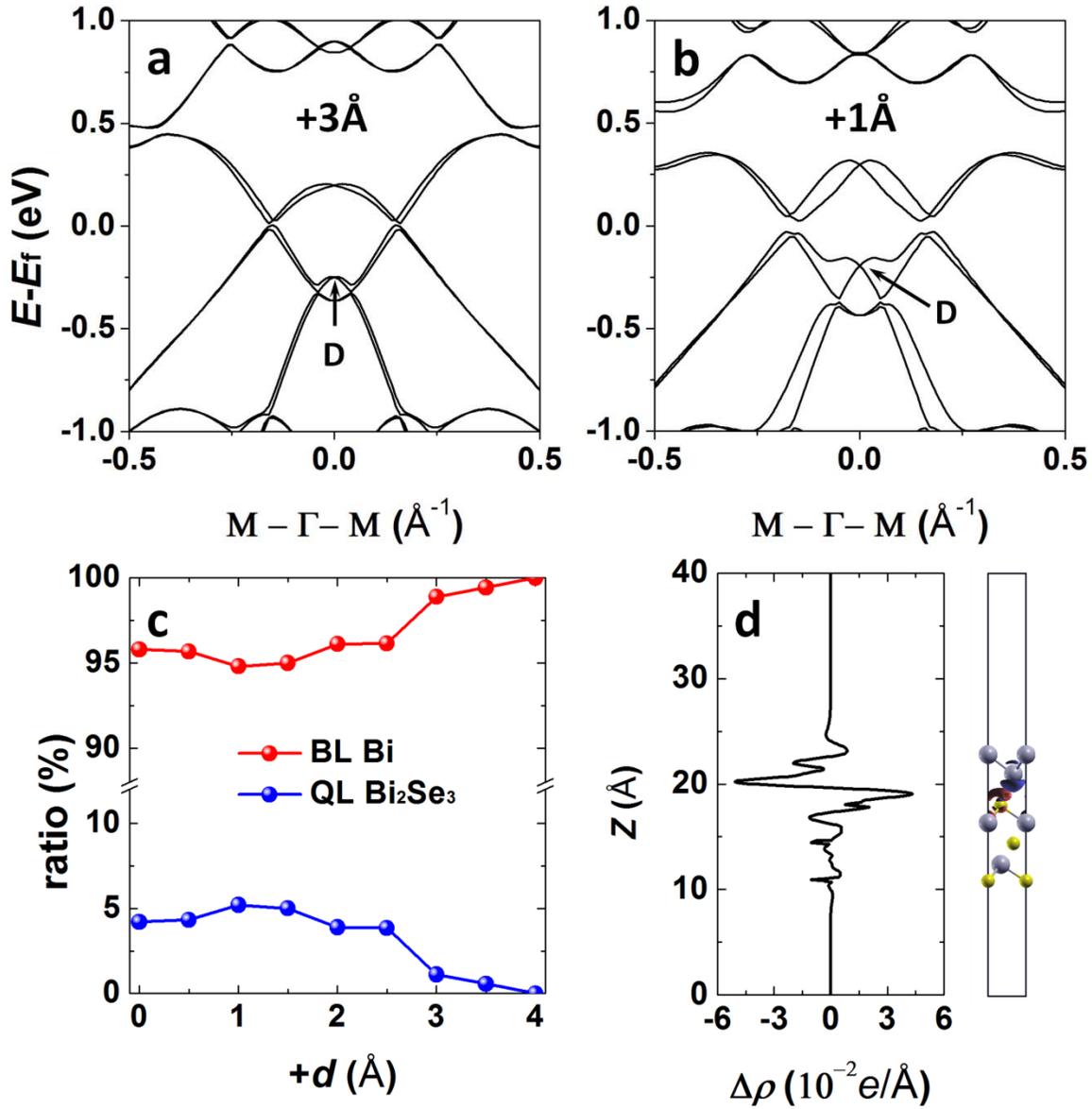

**Figure 4. Model DFT bands as a function of interfacial distance.** (a) and (b) Bands at the artificially increased interfacial distance by 3 and 1 Å, respectively, showing the creation process of helical Dirac cone. (c) Spectral weight at the Dirac point resolved between the Bi BL and $Bi_2Se_3$ QL as a function of interfacial distance. (d) Interfacial charge transfer between Bi BL and $Bi_2Se_3$ QL at the equilibrium distance, showing the internal electrical field.

18